\documentclass[12pt]{article}
\usepackage{pdproc,epsfig} 

\begin{document}

\renewcommand{\thefootnote}{\alph{footnote}}
  
\title{
DO NEUTRINOS VIOLATE CP?~\protect\footnote{
Written version of a talk presented at the 
``Third International Workshop on Neutrino Oscillations in Venice'' 
(NO-VE 2006), Venice, Italy. 7-10, February 2006.}} 

\author{HISAKAZU MINAKATA}

\address{Department of Physics, Tokyo Metropolitan University, \\
1-1 Minami-Osawa, Hachioji, Tokyo 192-0397, Japan\\
 {\rm E-mail: minakata@phys.metro-u.ac.jp}}




\abstract{
In trying to answer the question in the title of my talk, 
I have argued on ground of naturalness that leptonic CP violation is 
very likely to exist both in the form of the Kobayashi-Maskawa type 
and the Majorana type phases. 
The latter part of the argument has to be backed up by a general 
argument by Yanagida which states that neutrinos must be Majorana 
particles because our universe is asymmetric with respect to 
baryon number. The argument is reviewed. 
Since the naturalness argument raises the possibility of naturally small 
$\theta_{13}$ and $\theta_{23} - \pi/4$, I discuss possible 
experimental methods for probing into these two small quantities. 
They include recent proposal of the resonant $\bar{\nu}_{e}$ absorption reaction 
enhanced by the M\"ossbauer effect which may allow extremely high 
sensitivity for not only $\theta_{13}$ but also $\Delta m^2_{31}$. 
The issue of how to resolve the $\theta_{23}$ octant degeneracy is briefly 
discussed with emphasis on the atmospheric neutrino observation and the 
reactor-accelerator combined methods. 
}
   
\normalsize\baselineskip=15pt

\section{It is a tough question, isn't it?}

I am happy to be here again, the unique ``aqua city'', under 
the kind invitation by Milla, who was also so kind to give me such 
a tough question as in the title of my talk{\em !} 
But, since it is my duty to give an answer to her question, 
let me start my lecture from a trial of answering it. 
I don't know how far I can go, but it is nice if you find some of the 
comments below enjoyable to you. 

Clearly, observing leptonic CP violation is one of the prime targets 
of the future neutrino experiments. 
It is so because CP violation is one of the unresolved mystery in 
particle physics, and people feel something deep in that. Furthermore, 
since we now have the successful description of CP violation in 
the quark sector, the Kobayashi-Maskawa mechanism \cite{KM}, 
understanding of leptonic CP violation will 
shed light to the lepton-quark correspondence \cite{correspondence}. 
Even more interestingly, the lepton sector might have another 
source of CP violation thanks to the possible Majorana nature 
of neutrinos. It is likely that this question is related to another 
intriguing question of baryon number asymmetry in the universe 
\cite{leptogenesis}, as we will see below.

\section{Do neutrinos violate CP?; Kobayashi-Maskawa type CP violation}

Let us start by asking if there is any CP violation in the lepton sector 
due to the Kobayashi-Maskawa type phase $\delta$ in the lepton 
flavor mixing matrix, the MNS matrix \cite{MNS}. 
Since it is a complete analogue of the CKM matrix \cite{KM,cabibbo} 
in the quark sector, we can consult to many particle theory 
textbooks on how it is defined. They say, 
\begin{eqnarray}
U_{MNS} = S^{\dagger}(l) S(\nu)
\label {MNS}
\end{eqnarray}
where $S(\nu)$ and $S(l)$ denote unitary matrices which diagonalize 
the mass matrices of neutrinos and the charged leptons, respectively. 
The phase which is responsible for leptonic CP violation comes from 
physics of neutrino or charged lepton masses, or from both.  
Let us assume for definiteness that the neutrinos are Dirac particles.
Then, the each unitary matrix $S(\nu)$ and $S(l)$ has two phases 
even after their left phases absorbed into the wave functions.

The question is: 
Is there any chance that these phases all cancel out when we take 
combination $S^{\dagger}(l) S(\nu)$? 
We argue that it is highly unlikely. 
If you actually compute $S^{\dagger}(l) S(\nu)$ by assuming for 
each of $S(l)$ and $S(\nu)$ 
the standard form of the CKM matrix with extra right phases,  
you will be convinced of how unlikely is the cancellation.   
The values of phases of two matrices $S(\nu)$ and $S(l)$ have 
to be arranged so that they precisely cancel with each other when 
they meet after experiencing shifts due to non-Abelian nature of the 
building block of the MNS matrix.

I have another argument against the possible cancellation of 
CP violating phases in the MNS matrix.
To indicate the point, I propose to compare the two things, 
accidentally small $\theta_{13}$ and accidentally small $\delta$.
Can they be equally natural?

I answer the question in the negative for symmetry reasons. 
People invented some symmetries \cite{symmetry}, most of whose 
may have rooted in more phenomenological 
$\mu \leftrightarrow \tau$ exchange symmetries \cite{mu-tau} 
motivated by the nearly maximal atmospheric angle $\theta_{23}$. 
(I have no idea if the lists in \cite{symmetry} and \cite{mu-tau} are 
complete, because of so many references; 
Apology to any omission of relevant ones.)
The symmetries imply $\theta_{23}=\pi/4$ and $\theta_{13}=0$ 
in the symmetry limit. 
Therefore, the small $\theta_{13}$ is {\em natural} according to 
the definition of naturalness \`a la t' Hooft \cite{naturalness}. 
On the other hand, no symmetry is known for naturally small $\delta$.
Therefore, absence of CP violation in the lepton sector 
due to the Kobayashi-Maskawa type phase highly unlikely.

\section{Do neutrinos violate CP?; Majorana-type CP violation}

Most likely, the neutrinos are Majorana particles as preferred by 
a variety of models, most notably by the see-saw mechanism \cite{seesaw}. 
If it is the case, we will have CP violation due to extra Majorana 
phases in the lepton flavor mixing matrix \cite{Mphase}. 
Since there are two Majorana phases in three-generation neutrinos, 
the possibility of accidentally small CP violation is even more unlikely 
in the case of Majorana-type CP violation.

Therefore, the real question is; Are the neutrinos Majorana particles? 
There is a strong argument given by Yanagida \cite{yanagida} 
which answers in the positive to this question. 
So let me introduce it for you, assuming that it is 
not familiar to the audience. 
His argument starts from the well known facts on which everybody would agree:

\begin{itemize}

\item
We know that our universe is asymmetric with respect to baryon number. 

\item
We know that above $\sim$1 TeV the only meaningful quantum number is 
$B-L$, not the baryon number $B$ or the lepton number $L$ separately, 
because of the anomaly in the Standard Model \cite{tHooft}, or more 
precisely speaking, the gsphaleronh effect \cite{KRS}.\footnote{
Here is some comments on the gsphaleronh for those who are not familiar to it. 
Everybody knows that because of the instanton configuration the gauge 
theory vacuum is enriched by the periodic vacua which differ by the topological 
winding number and are separated by a barrier whose hight is given by 
$\sim M_{W}/\alpha$. 
(Consult, e.g., to Coleman's lectures \cite{coleman} for more about it.)  
The sphaleron is nothing but the field configuration at the top of the barrier 
\cite{manton}. 
One can show that by tunneling to the adjacent vacuum the fermion number 
(baryon or lepton number) changes by one unit due to the  anomaly in 
chiral gauge theories. 
The transition conserves $B-L$ because it is anomaly free.
Now at zero temperature the transition is severely suppressed by the 
penetration factor \cite{tHooft}. 
But, Kuzmin, Rubakov, and Shaposhnikov \cite{KRS} pointed out that 
the transition can proceed at high temperature $T$, and have shown that 
the rate at temperature around the electroweak phase transition 
can be calculable by using sphaleron configuration. 
It is natural because $M_{sphaleron}/T$ characterize the difficulty 
or easiness of transition taking place due to thermal effects.
A fair computation exists to support the conclusion \cite {mclerran}. 
Thus, all the nonvanishing $B-L$ generated in earlier cosmological 
evolution is expected to be wiped out at the time of electroweak transition.
}

\item
Therefore, we must have $B+L$ generation in some stages of the 
cosmological evolution to have nonzero baryon number to date. 

\end{itemize}

I hope that all of them above are agreeable by everybody. 
(I asked the audience in the lecture room if someone disagrees with 
any of the statements above, but no one did.) 
If so, here is the second step in the Yanagida argument:

\begin{itemize}

\item
Let us assume the Standard Model of particle physics. 
Then, there is no renormalizable operator which violates $B-L$, 
and hence there is no chance of generating baryon number asymmetry 
(unless it is so carefully designed as to evade the sphaleron extinction). 
Therefore, we must go beyond the Standard Model to have nonzero 
baryon number in the universe. 

\item

The model independent way of searching for the possible $B-L$ 
violating operator is to look for suitable higher dimensional operators 
\cite{higher-dim}.
The unique lowest dimension operator which violates $B-L$ is 
\begin{eqnarray}
\frac{1}{M} \phi  \phi  \nu \nu 
\label {higher-dim}
\end{eqnarray}
which must exists so that baryon number asymmetry (and we ourselves) 
exists. 
Therefore, the Majorana mass term must exist for neutrinos. 

\end{itemize}

Two immediate comments are in order:

\noindent
(1) The formula in (\ref{higher-dim}) is nothing but the seesaw 
formula for neutrino masses \cite{seesaw}. 
In the present discussion, however, it is derived in 
a ``model-independent'' way.

\noindent
(2) I note that most likely the operator in (\ref{higher-dim}) is 
responsible for the neutrino mass observed by Super-Kamiokande 
\cite{SKatm} and KamLAND \cite{KamLAND} 
experiments.\footnote{
The significance of the KamLAND experiment which established 
the mass-induced neutrino oscillation as the unique interpretation 
of the solar neutrino flavor transformation has been emphasized 
by Valle in his talk in this workshop \cite{valle}. 
}  
The  former atmospheric oscillation was confirmed by K2K \cite{K2K}, and 
the latter solar oscillation has been hinted by the long-term extensive 
efforts by various solar neutrino observation \cite{solar} which was initiated 
by the pioneering Davis experiment \cite{davis} and 
has been concluded by SNO \cite{SNO}.

\section{Naturally small $\theta_{13}$ and/or $\theta_{23} - \pi/4$ ?}

I argued above, on ground of naturalness, that the Kobayashi-Maskawa 
type CP violating phase $\delta$ is unlikely to be canceled between 
the two unitary matrices which diagonalize the neutrino and the 
charged lepton mass matrices. 
The argument raises the possibility that $\theta_{13}$ could be tuned 
to be very small and at the same time $\theta_{23}$ be close to the 
maximal. 
Therefore, I would like to address these issues in the rest of my talk.

Since $\mu \leftrightarrow \tau$ exchange symmetry is badly broken 
(note that $m_{\tau} \simeq 20~m_{\mu}$), the predictions 
$\theta_{13}=0$ and $\theta_{23}= \pi/4$ cannot be exact. 
It is important to try to compute deviations of the results obtained 
in the symmetry limit. 
Only by finding correlation between these two small quantities 
one can establish the symmetry, if any, by distinguishing it from 
some other possibilities such as the quark-lepton complementarity 
\cite{QLC} extended to 2-3 sector which also suggests that 
$\theta_{23}$ is close to $\pi/4$. 
At the moment, however, 
we do not have a reliable theoretical machinery to compute them.

I focus here on the possible experimental 
methods for determining the small corrections to the symmetry limit.
However, you may ask the question; 
The method for measurement of $\theta_{13}$ has been extensively 
discussed by using varying experimental means; 
accelerator \cite{JPARC,NOVA,OPERA}
reactor \cite {krasnoyarsk,MSYIS,reactor_white}, and 
astrophysical neutrinos \cite{astro13}. 
Are there any other possibilities to explore? 
Amazingly, the answer seems {\em Yes}.

\section{M\"ossbauer enhanced resonant absorption of monochromatic 
antineutrino beam}

Recently, it was proposed \cite{raghavan} that the 
the resonant absorption reaction \cite{mikaelyan}
\begin{eqnarray}
\bar{\nu_{e}} + ^{3}\mbox{He} + \mbox{orbital e}^{-} \rightarrow ^{3}\mbox{H}
\label{res-abs}
\end{eqnarray}
with simultaneous capture of an atomic orbital electron can be 
dramatically enhanced by using the inverse reaction 
$^{3}\mbox{H}  \rightarrow \bar{\nu_{e}}+$$^{3}\mbox{He} + \mbox{orbital e}^{-}$, 
by which the resonance condition is automatically satisfied. 
(See \cite{visscher,schiffer} for earlier suggestions.)  
Furthermore, by embedding both the source $^{3}\mbox{H}$ and 
the target $^{3}\mbox{H}$ atoms into solid the overlap between the line 
widths of the emission and the absorption can be dramatically 
improved, which may lead to the enhancement of the reaction 
cross section of (\ref{res-abs}) by a factor of $\sim 10^{11}$ \cite{raghavan}. 
To realize the enhancement it is important to secure that 
both the source and the target atoms are placed in a metal so that 
they can enjoy the same environment.\footnote{
If the line shift occurs between the source and the target atoms 
it may be cancelled by gravitational effect by placing them 
in a different elevation. 
But, since we want to remain in a suitable underground site, 
the hight difference between the source and the target would practically 
be less than $\sim$100 m ($\sim$10 m for the $\theta_{13}$ experiment). 
This places a limit on absolute value of the relative line shift 
manageable by this method to the order of 
$< 2 \times 10^{-10}$ ($2 \times 10^{-11}$) eV.
}

One might naively think that the probability of having beta decay 
with simultaneous capture of electrons into the atomic orbit is tiny. 
But, the author of \cite{bahcall} argues that the process occurs 
not by capturing an emitted electron to the orbit but by {\em creating} 
an electron into the orbit. In fact, the calculated branching fraction 
of the bound state beta decay to the conventional electron emitting 
decay is not so small, $4.7 \times 10^{-3}$. 
Therefore, it appears that the possibility deserves further attention 
which, I hope, would trigger closer examinations of its experimental 
feasibility.

\subsection{10 m baseline $\theta_{13}$ experiments}

Why the new proposal interesting in the context of measurement 
of small $\theta_{13}$? 
First of all, the ultra-low neutrino energy of 18.6 keV of (\ref{res-abs}) 
makes it possible, with the first oscillation maximum of 
$L_{\mbox{\mbox{OM}} } = 
9.2~(\Delta m^2_{31} / 2.5 \times 10^{-3} \mbox{eV}^2)^{-1}$ m, 
to design a 10 m baseline $\theta_{13}$ experiment. 
Furthermore, the M\"ossbauer enhancement 
of the reaction cross section of (\ref{res-abs}) to 
$\sigma_{res} \simeq 5 \times 10^{-32} \mbox{cm}^2$ 
enables us an enormous statistics 
\begin{eqnarray}
R_{\mbox{enhanced}} = 1.2 \times 10^{6}
\left(  \frac{S M_{T}}{1 \mbox{MCi} \cdot \mbox{100~g} }  \right) 
\left(  \frac{L}{10~\mbox{m} }  \right)^{-2} 
\mbox{day}^{-1}, 
\label{enhancedR}
\end{eqnarray}
a million events a day by using 100 g ({\em not} 100 kiloton!) of 
$^3$He target, assuming 1 MCi source.

We have argued in \cite{mina-uchi} that if the direct counting of 
produced $^3\mbox{H}$ atom works, the relative systematic error can be 
as low as 0.2\% by a movable detector setting, 
and if not it may be of the order of $\simeq$1\%. 
For concreteness, we restrict ourselves here to a particular setting 
described as Run IIB in \cite{mina-uchi}: 

\vspace{0.2cm}
\noindent 
Run IIB: Measurement at 10 different detector locations; 
$L= L_{i}$ ($i=1, ... 10$) where $L_{i+1} = L_{i} + \frac{2}{5} L_{\mbox{OM}} $ and 
$L_{1} = \frac{1}{5} L_{\mbox{OM}} $  
so that the entire period, 
$\Delta = 0$ to 2$\pi$, is covered. At each location an equal number of 
$10^{6}$ events is to be collected. 

\vspace{0.3cm}

\noindent 
The huge statistics and the controlled uncorrelated systematic error 
to 0.2\% (1\%) level should allow precision measurement of 
$\theta_{13}$ up to $\sin^2 2\theta_{13} = 0.002$ (0.008) or so 
at 1$\sigma$ CL \cite {mina-uchi}.

\subsection{Possible extreme accuracy in $\Delta m^2_{31}$ measurement}

Though slightly off line from the present discussion, it is worth to 
mention the additional physics potential of 10 m $\theta_{13}$ 
experiment. 
The $\bar{\nu}_{e}$ beam from the tritium decay is monochromatic 
for practically all purposes even before the M\"ossbauer enhancement. 
Then, it is natural to think about precision measurement of 
$\Delta m^2_{31}$ by using the recoilless resonant absorption. 
This expectation was confirmed in \cite{mina-uchi} in which the accuracy 
of $\Delta m^2_{31}$ determination is shown to reach sub percent level.

\begin{figure}[h]
\vspace*{1cm}
\begin{center}
\epsfig{figure=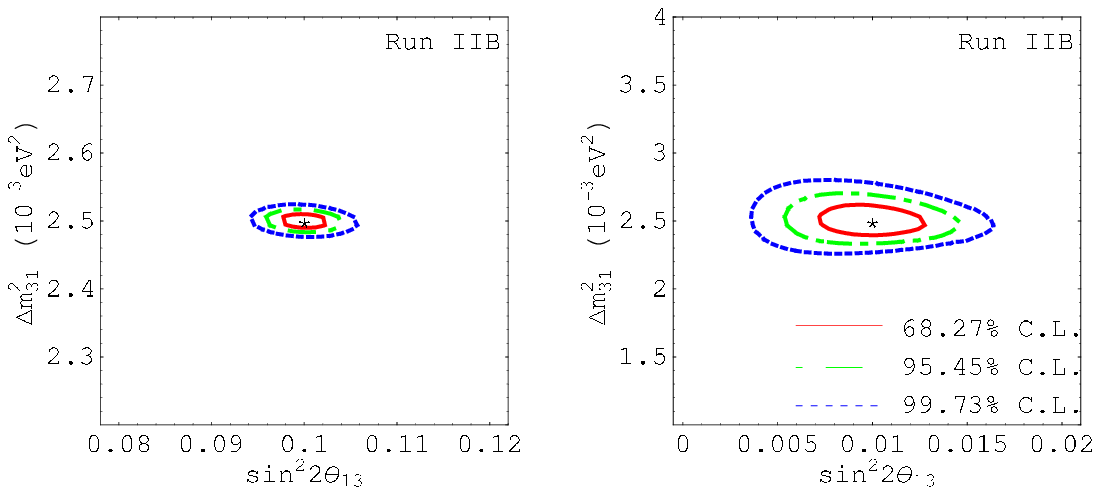,width=13cm}
\end{center}
\vspace*{0.5cm}
\caption{
The expected allowed region by Run IIB defined in the text with 
number of events $10^{6}$ in each location are depicted. 
The red-solid, the green-dashed, and the 
blue-dotted lines are for 1$\sigma$ (68.27\%),  
2$\sigma$ (95.45\%), and 3$\sigma$ (99.73\%) CL for 2 DOF, respectively. 
The input values of the mixing parameters are marked by asterisks and 
they are as follows: 
$\Delta m^2_{31} = 2.5 \times 10^{-3}$ eV$^2$, 
$\sin^2 2\theta_{13}=0.1$ and 0.01 
in the left and the right panels. 
}
\label{IIB}
\end{figure}

For an exposure of Run IIB defined above, the allowed regions at 
1$\sigma$-3$\sigma$ CL are given in Fig.~\ref{IIB}. 
The figures are for the uncorrelated systematic error of 0.2\%.

\vspace*{0.5cm}

By optimizing on $\theta_{13}$, we have obtained for 1 DOF 
the following sensitivity to $\Delta m^2_{31}$ with Run IIB; 
If the uncorrelated systematic error of 0.2\% is realized, the accuracy 
of measurement of $\Delta m^2_{31}$ is 
$\simeq 0.3~(\sin^2 2\theta_{13} / 0.1)^{-1}$\% at 1$\sigma$ CL. 
For the pessimistic systematic error of 1\% the sensitivity is worsen by 
about a factor of four. 
For details of the analysis procedure and results for various settings, 
see \cite{mina-uchi}.

The obvious possible application of the extreme sensitivity to 
$\Delta m^2_{31}$ is the method for determining the neutrino 
mass hierarchy by comparing two kind of disappearance measurement 
$\nu_{e} \rightarrow \nu_{e}$ and $\nu_{\mu} \rightarrow \nu_{\mu}$, as 
proposed in \cite{NPZ,GJK}.

\section{Which octant does $\theta_{23}$ live?}

Determining which octant $\theta_{23}$ lives and how far it is 
from the maximal angle $\pi/4$ is not an easy question to answer.
Nevertheless, it is important to find ways to solve it because 
the $\theta_{23}$ octant degeneracy \cite{octant} is one of the major 
obstacle in precision determination of $\theta_{23}$ \cite{MSS04}.

Principle of resolving the $\theta_{23}$ degeneracy is simple; 
Look for oscillation channels which depend upon $\theta_{23}$ 
not through the combination $s^2_{23} \sin^2 2\theta_{13}$. 
However, it can be shown in mostly by analytic manner that 
it is very difficult to resolve the $\theta_{23}$ degeneracy only 
by accelerator experiments with modest baseline of 
$L < 1000$ km \cite {resolve23}.\footnote{
If it would be possible to create a very long baseline experiment with e.g., 
$L=6000$ km, there could be ways to circumvent the argument. 
But, it is {\em hard} to create intense enough beam or build huge 
detectors which can compensate the flux depletion proportional to 
$L^{-2}$, and to prepare beam line pointing toward them well 
below the horizon. 
}
Thus, at the moment there are two ways, to my knowledge, to resolve 
the $\theta_{23}$ degeneracy. 
Let us discuss them briefly one by one.

\subsection{High statistics atmospheric neutrino observation}

\begin{figure}[h]
\vspace*{0.5cm}
\begin{center}
\epsfig{figure=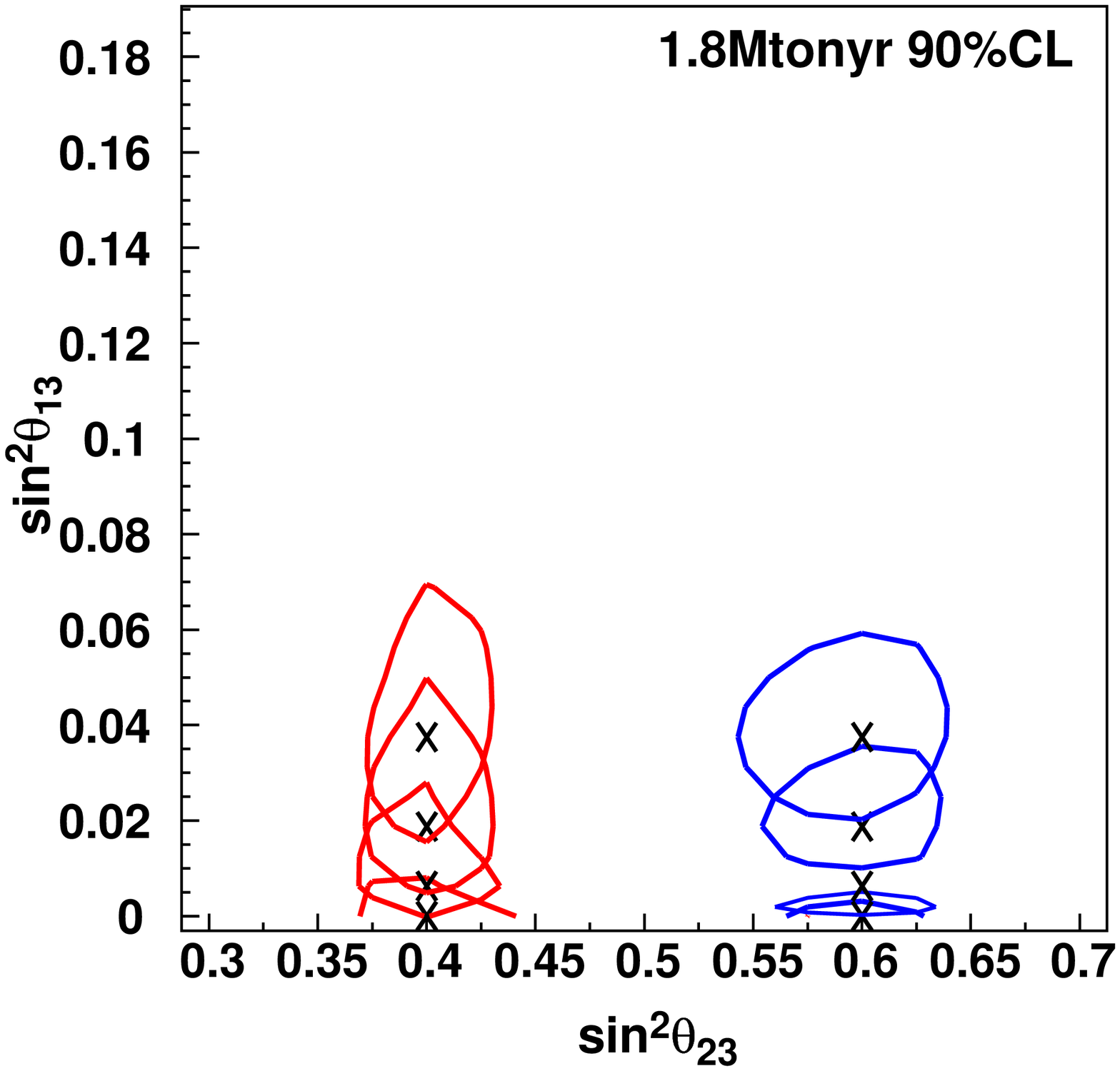,width=7cm}
\epsfig{figure=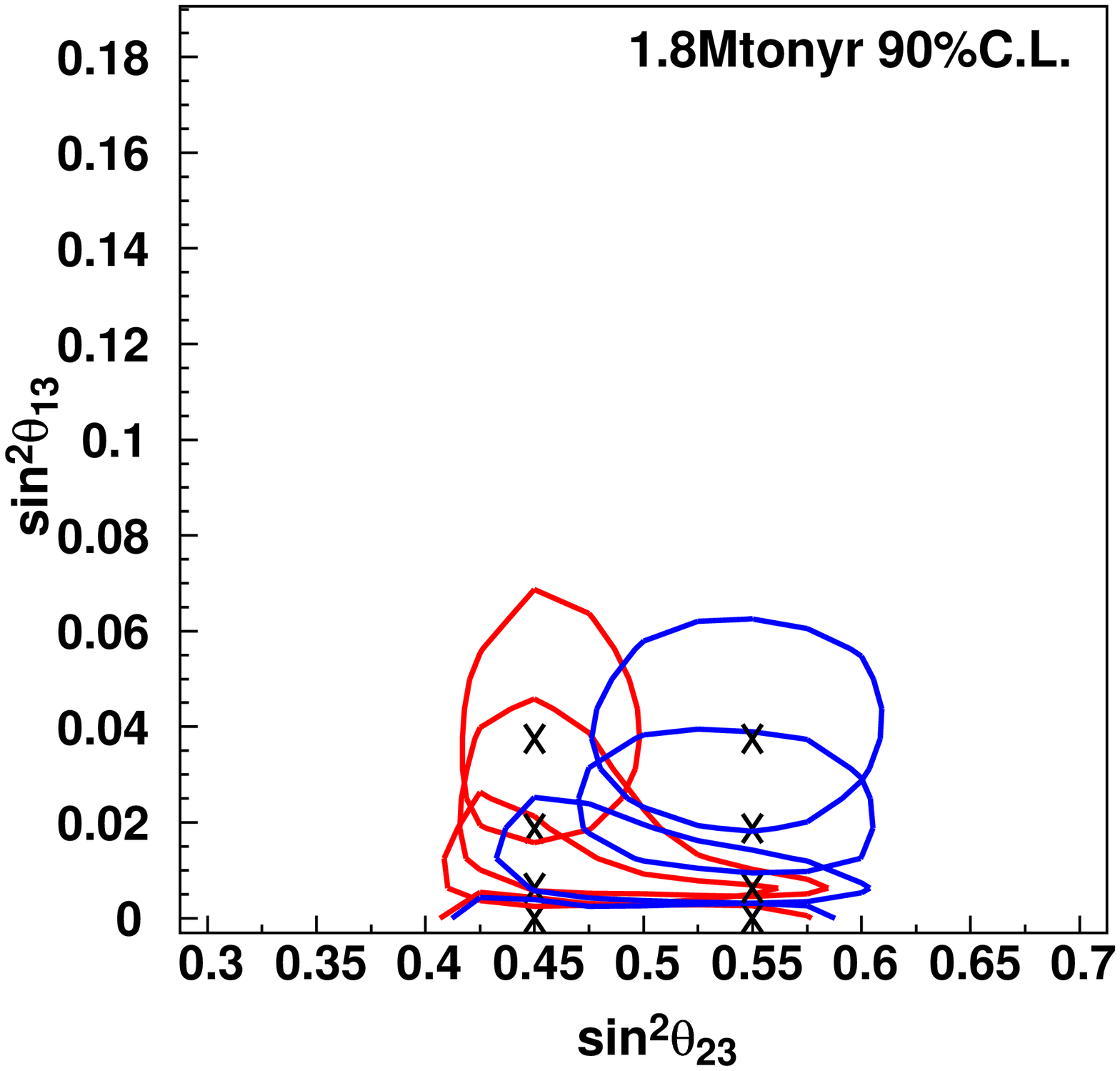,width=7cm}
\end{center}
\vspace*{0.3cm}
\caption{
The discrimination between the first and the second octant of 
$\theta_{23}$ by 1.8 Mton$\cdot$year (3.3 years of HK) 
exposure of atmospheric neutrinos. 
The CP phase is taken as $\delta=\pi/4$. 
Discrimination can be done for relatively large deviation of $\theta_{23}$ 
from the maximal down to a vanishingly small $\theta_{13}$ (left figure). 
On the other hand, it is getting very hard for a smaller deviation, 
$\sin^2 2\theta_{23}=0.99$ (right figure). 
Courtesy by Takaaki Kajita, also presentatd at 
Next Generation of Nucleon Decay and 
Neutrino Detectors (NNN05) \protect\cite{atm23}. 
}
\label{atm23}
\end{figure}

The atmospheric method for resolving the $\theta_{23}$ degeneracy 
utilizes the solar oscillation term which is proportional to 
$c^2_{23}$ and independent of $\theta_{13}$ in a good approximation. 
Therefore, its sensitivity to the $\theta_{23}$ degeneracy essentially 
relies on detection capability of the solar term \cite{concha-smi_23,choubey2}. 
Since the term is independent of $\theta_{13}$ the method works 
even for vanishingly small $\theta_{13}$. 
See Fig.~\ref{atm23} which is taken from \cite{atm23}.
On the other hand, it requires enormous statistics which requires 
the current Super-Kamiokande to run $\sim$80 years.
Clearly, construction of much larger detector such as 
Hyper-Kamiokande is {\em the necessity}.
%

\subsection{Reactor-accelerator combined method}

The other possibility of resolving the $\theta_{23}$ degeneracy is to 
combine reactor measurement of $\theta_{13}$ to accelerator 
$\nu_{\mu}$ disappearance and $\nu_{e}$ appearance experiments 
\cite{MSYIS}. (See \cite{octant} for earlier suggestion.) 
The principle is very simple; 
The accelerator disappearance and appearance measurement determine 
$\sin^2 2\theta_{23}$ and $s^2_{23} \sin^2 2\theta_{13}$, respectively, 
leaving a degenerate solution if $\theta_{23}$ is not maximal. 
The reactor measurement of $\theta_{13}$, which is largely 
independent of other mixing angles, picks up one of the solutions.

\begin{figure}[h]
\vspace*{0.2cm}
\begin{center}
\epsfig{figure=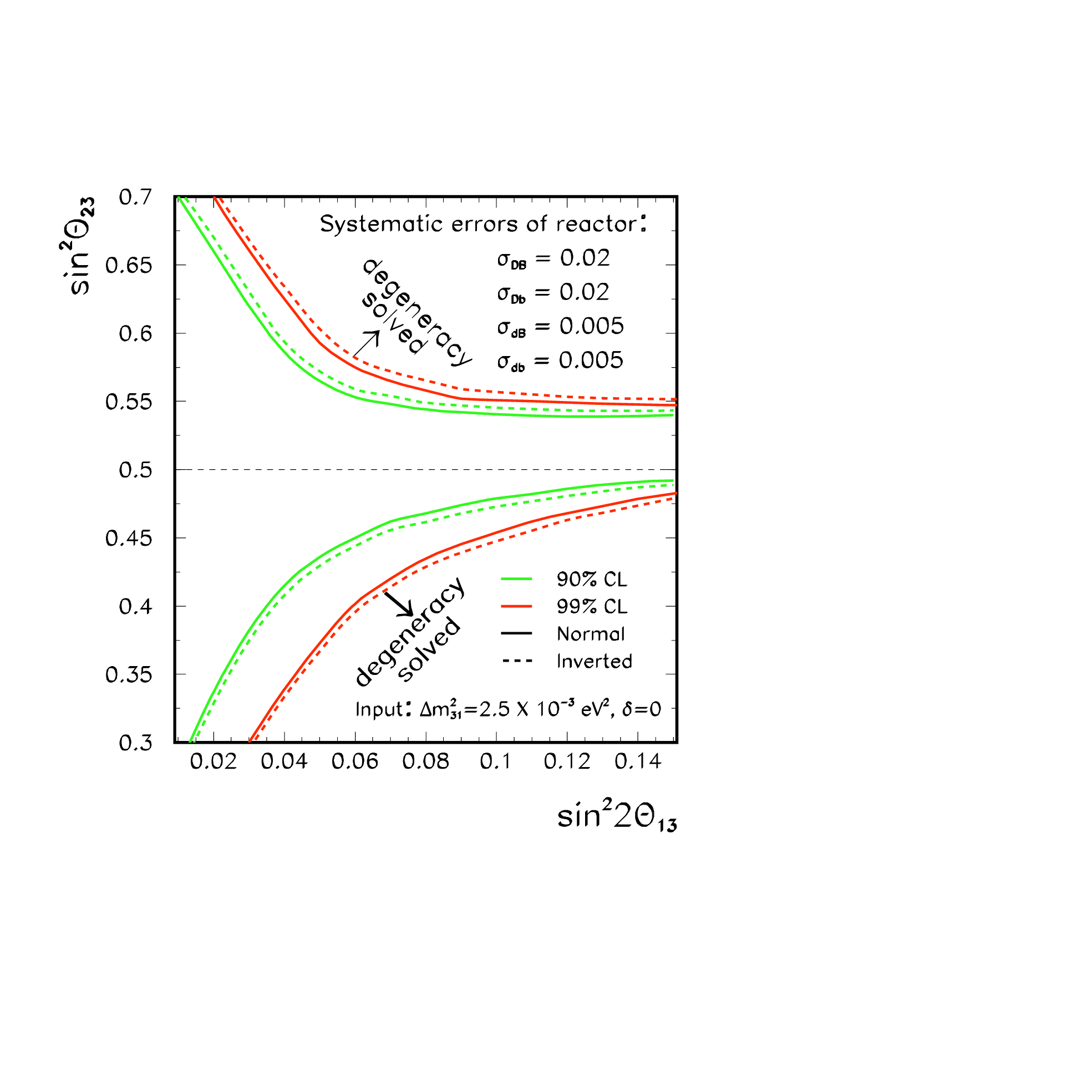,width=9cm}
\end{center}
\caption{
The region in $\sin^2 2\theta_{13} - \sin^2 \theta_{23}$ space 
where the $\theta_{23}$ octant degeneracy can be resolved 
at 90\% (thin green) and 99\% (thick red) CL. 
The solid (dashed) curve is for the case taking the normal (inverted) 
hierarchy while performing the fit, assuming the normal hierarchy 
as an input.  
The conservative systematic errors, as indicated in the figure, are
considered here.
}
\label{region_resolved1}
\end{figure}

\begin{figure}[h]
\vspace*{0.2cm}
\begin{center}
\epsfig{figure=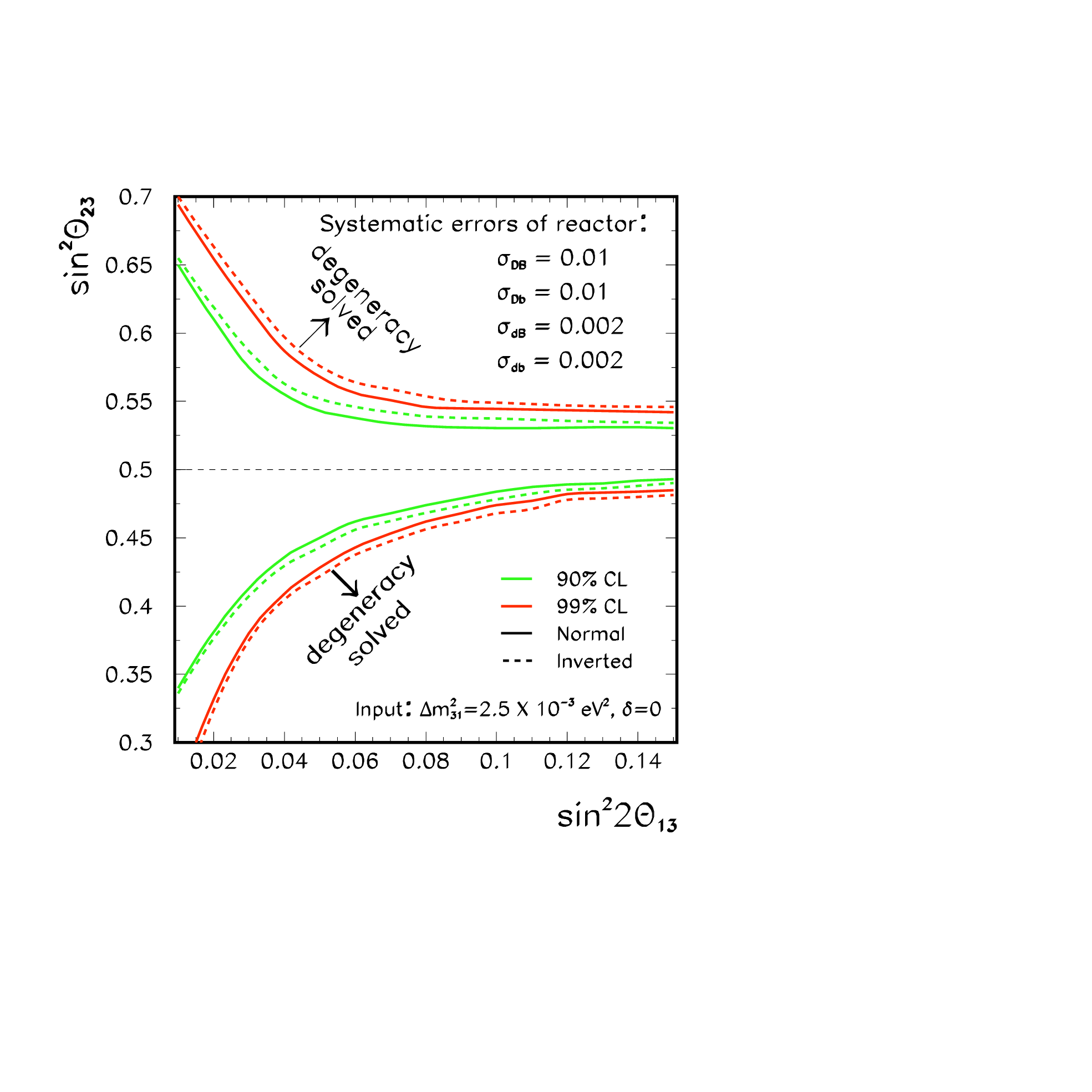,width=9cm}
\end{center}
\caption{
The same as in Fig.~\ref{region_resolved1} but with the 
optimistic systematic errors. 
}
\label{region_resolved2}
\end{figure}

Quite recently, we have revisited the idea to examine quantitatively 
the limit of resolving power of the $\theta_{23}$ degeneracy by this 
method \cite{resolve23}. 
We have assumed for accelerator experiment the phase II of the T2K 
experiment with 2 (6) years running of neutrino
(anti-neutrino) modes with 4MW beam power with the Hyper-Kamiokande
detector whose fiducial volume is 0.54 Mt \cite{JPARC}.  For the
reactor experiment, the exposure of 10 GW$\cdot$kt$\cdot$yr is
assumed.
For the T2K II experiment the systematic errors are assumed to be 2\%.  
Since the accuracy of the reactor measurement of $\theta_{13}$ 
is of crucial importance for the sensitivity of resolving the degeneracy 
we have examined two sets of the systematic errors: \\
Pessimistic errors; 
detector correlated errors of 2\% and uncorrelated errors of 0.5\%. \\
Optimistic errors; 
detector correlated errors of 1\% and uncorrelated errors of 0.2\%. \\
(See \cite{resolve23} for details of the systematic errors.)  

The resultant sensitivity regions obtained by assuming 
the pessimistic and the optimistic systematic errors are given in 
Figs.~\ref{region_resolved1} and Figs.~\ref{region_resolved2}, 
respectively. 
At relatively large $\theta_{13}$ the method is shown to be powerful 
in resolving the $\theta_{23}$ degeneracy. 
At small  $\theta_{13}$, however, resolving power of the degeneracy is 
limited even for the case of optimistic systematic errors. 
It is notable that resolving power 
of the degeneracy is not symmetric with respect to $\theta_{23}=\pi/4$; 
It is easier to resolve the degeneracy for $\theta_{23}$ in the 
first (second) octant for relatively large (small) $\theta_{13}$. 
It appears that it is the result of intricate interplay of the various 
factors \cite{resolve23}.

I note that at large $\theta_{13}$ in particular in the first octant the 
reactor-accelerator method has better sensitivities, while the 
atmospheric method wins at small  $\theta_{13}$. 
To improve the resolving power of the former we need a better 
accuracy in $\theta_{13}$ determination. 
Naturally, the 10 m baseline experiment using the 
M\"ossbauer enhanced resonant absorption of 
monochromatic antineutrino beam discussed in the previous 
section might be of help.

\section{Miscellaneous remarks}

My presentation in the workshop included remarks on miscellaneous 
topics including 
(1) introducing the bi-probability plot \cite{MNjhep01} for intuitive 
understanding of the CP phase-matter effect interplay, 
(2) importance of use of two-detector setup to detect CP violation 
\cite{MNplb97}, 
(3) some comments on how to solve parameter degeneracies.

\begin{figure}[h]
\begin{center}
\epsfig{figure=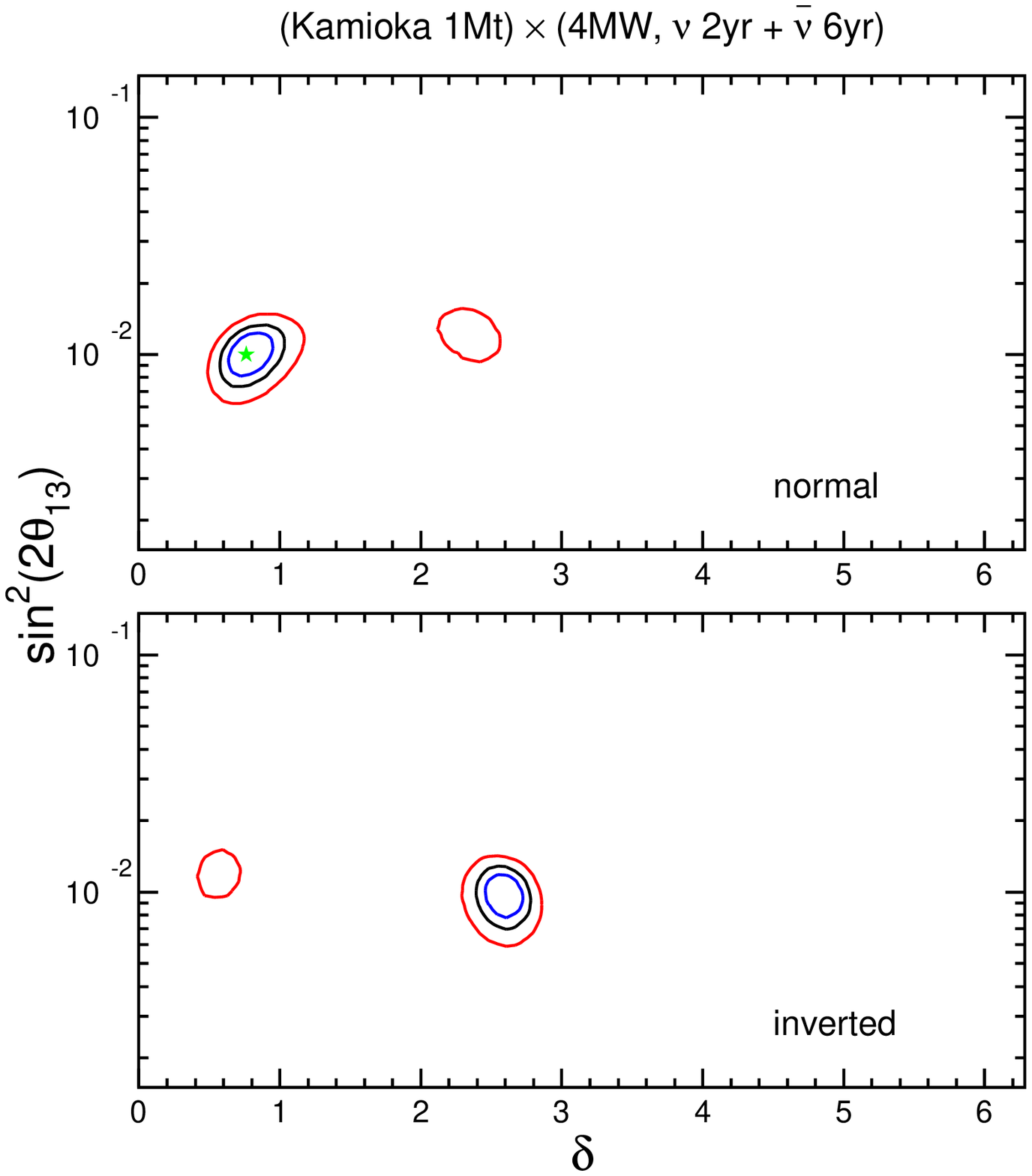,width=10cm}
\end{center}
\vspace*{0.2cm}
\caption{
The region allowed in $\delta-\sin^2 2\theta_{13}$ space 
by 2 years of neutrino and 6 years of antineutrino running in the 
T2K II experiment. Taken from supplementary figures behind the 
reference \protect\cite{T2KK}. 
The true solutions are assumed to be located at 
($\sin^2{2\theta_{13}}$ and $\delta$) = (0.01, $\pi/4$) 
with positive sign of $\Delta m_{31}^2$, as indicated as the green star. 
Three contours in each figure correspond to
the 68\% (blue line), 90\% (black line) and 99\% 
(red line) C.L. sensitivities, which are defined
as the difference of the $\chi^2$ being 2.30, 4.61 and 9.21, respectively.
}
\label{intrinsic}
\end{figure}

In particular, I emphasized the role of spectrum analysis in resolving the 
so called intrinsic degeneracy \cite{intrinsic}. 
It was my prejudice that the intrinsic degeneracy is hard to resolve 
because the differences between the two degenerate solutions, 
$\theta_{13}$ and $\sin \delta$, are so tiny \cite{MNP2}. 
However, we have learned in exploration of the idea of the 
Kamioka-Korea two identical detector complex (T2KK) 
that it is the easiest degeneracy to lift. 
(For T2KK itself see \cite{T2KK} and \cite{kajita}.) 
In Fig.~\ref{intrinsic}, which is just one of thousand figures behind 
the paper \cite{T2KK}, it is illustrated that the spectrum analysis 
by HK placed in Kamioka only (without a Korean detector) is powerful 
enough to (almost) resolve the intrinsic degeneracy despite a rather 
small value of $\theta_{13}$, $\sin^2 2\theta_{13}=0.01$. 

Notice, however, that this setting does {\em not} resolve the degeneracy 
caused by the unknown mass hierarchy \cite{MNjhep01} at all. 
To resolve both of the degeneracy simultaneously we need T2KK. 
It is also notable that the degenerate solutions have ``X-shaped'' 
structure which can be understood as a consequence of 
cooperation of a symmetry behind the sign-$\Delta m^2$ 
degeneracy \cite{MNjhep01} and the property of the intrinsic degeneracy.

\section{Conclusion}

I have concluded my talk with several short remarks:

\begin{itemize}

\item

Leptonic CP violation, due to both the Kobayashi-Maskawa type 
and the Majorana-type phases is very likely to exist. 

\item

There are still rooms (referring to T2KK and other ideas) to make 
progress along the line of conventional superbeam \cite{superbeam} 
to explore leptonic CP violation. 

\item

New opportunities seem exist in physics to be done with the 
M\"ossbauer enhanced resonant absorption of 
monochromatic neutrino beam.  

\item

Among the parameter degeneracies the $\theta_{23}$ octant 
degeneracy may be the hardest one to solve.  
New ideas and/or gigantic detectors are called for.

\end{itemize}

\section{Acknowledgements}

I thank Tsutomu Yanagida for useful correspondences. 
I am grateful to all the collaborators of the works 
\cite{T2KK,resolve23,mina-uchi} for enjoyable collaborations 
through which my recognitions to the topics discussed in this 
manuscript have been deepened. 
I thank Departamento de F\'{\i}sica, 
Pontif{\'\i}cia Universidade Cat{\'o}lica (PUC) do Rio de Janeiro 
for hospitality where this manuscript was written. 
My visit is supported by Bilateral Programs, Scientist Exchanges based on 
Japan Society of Promotion of Science and Brazilian Academy of Sciences.
This work was supported in part by the Grant-in-Aid for Scientific Research, 
No. 16340078, Japan Society for the Promotion of Science.


\begin{thebibliography}{99}


\bibitem {KM}
M. Kobayashi and T. Maskawa, Prog. Theor. Phys. 
{\bf 49}, 652 (1973).

\bibitem {correspondence}
It varies from the  ``baroque'' version, 
A.~Gamba, R.~E.~Marshak and S.~Okubo, 
Proc. Nat. Acad. Sci. {\bf 45}, 881 (1959); 
Z.~Maki, M.~Nakagawa, Y.~Ohnuki, and S.~Sakata,
Prog. Theor. Phys.  {\bf 23}, 1174 (1960), 
%
to the modern one:
C.~Bouchiat, J.~Iliopoulos and P.~Meyer,
  Phys.\ Lett.\ B {\bf 38}, 519 (1972).


\bibitem {leptogenesis}
M.~Fukugita and T.~Yanagida,
  Phys.\ Lett.\ B {\bf 174}, 45 (1986).


\bibitem {MNS}
Z.~Maki, M.~Nakagawa and S.~Sakata,
Prog.\ Theor.\ Phys.\  {\bf 28}, 870 (1962).
See also, B.~Pontecorvo, 
Zh. Eksp. Teor. Fyz. {\bf 53}, 1717 (1967) 
[Sov. Phys. JETP {\bf 26}, 984 (1968)].  


\bibitem {cabibbo}
N. Cabbibo, Phys. Rev. Lett. {\bf 10}, 531 (1963). 


\bibitem{symmetry}
E.~Ma and G.~Rajasekaran,
  Phys.\ Rev.\ D {\bf 64}, 113012 (2001)
  [arXiv:hep-ph/0106291];
%
E.~ Ma, Mod.\ Phys.\ Lett.\ A {\bf 17}, 2361 (2003) 
[hep-ph/0211393];
%
K.~S.~Babu, E.~Ma, and J.~W.~F.~Valle, 
Phys.\ Lett.\ B {\bf 552}, 207 (2003) 
[hep-ph/0206292];
%
 W.~Grimus and L.~Lavoura,
  JHEP {\bf 0107}, 045 (2001)
  [arXiv:hep-ph/0105212]; 
%
Acta.\ Phys.\ Polon.\ {\bf B34}, 5393 (2003) 
[arXiv:hep-ph/0310050];
%
J.~Kubo, A.~Mondragon, M.~Mondragon and E.~Rodriguez-Jauregui,
Prog.\ Theor.\ Phys.\  {\bf 109}, 795 (2003)
[arXiv:hep-ph/0302196]; 
J.~Kubo,
Phys.\ Lett.\ B {\bf 578}, 156 (2004)
[arXiv:hep-ph/0309167];
%
K.~S.~Babu and J.~Kubo,
  Phys.\ Rev.\ D {\bf 71}, 056006 (2005)
  [arXiv:hep-ph/0411226];
%
W.~Grimus, A.~S.~Joshipura, S.~Kaneko, L.~Lavoura and M.~Tanimoto,
  JHEP {\bf 0407}, 078 (2004)
  [arXiv:hep-ph/0407112].
%
W.~Grimus, A.~S.~Joshipura, S.~Kaneko, L.~Lavoura, H.~Sawanaka and M.~Tanimoto,
  Nucl.\ Phys.\ B {\bf 713}, 151 (2005)
  [arXiv:hep-ph/0408123].
  
  
\bibitem{mu-tau}
 T.~Fukuyama and H.~Nishiura,
  arXiv:hep-ph/9702253; 
%
R.~N.~Mohapatra and S.~Nussinov,
  Phys.\ Rev.\ D {\bf 60}, 013002 (1999)
  [arXiv:hep-ph/9809415];
%
E.~Ma and M.~Raidal,
  Phys.\ Rev.\ Lett.\  {\bf 87}, 011802 (2001)
  [Erratum-ibid.\  {\bf 87}, 159901 (2001)]
  [arXiv:hep-ph/0102255]; 
 %
C.~S.~Lam,
  Phys.\ Lett.\ B {\bf 507}, 214 (2001)
  [arXiv:hep-ph/0104116]; 
%
P.~F.~Harrison and W.~G.~Scott,
  Phys.\ Lett.\ B {\bf 547}, 219 (2002)
  [arXiv:hep-ph/0210197]; 
%
Y.~Koide,
  Phys.\ Rev.\ D {\bf 69}, 093001 (2004)
  [arXiv:hep-ph/0312207]; 
%
T.~Kitabayashi and M.~Yasue,
  Phys.\ Rev.\ D {\bf 67}, 015006 (2003)
  [arXiv:hep-ph/0209294].
%
  Phys.\ Lett.\ B {\bf 621}, 133 (2005)
  [arXiv:hep-ph/0504212].
%
R.~N.~Mohapatra,
  JHEP {\bf 0410}, 027 (2004)
  [arXiv:hep-ph/0408187];
%
R.~N.~Mohapatra and W.~Rodejohann,
  Phys.\ Rev.\ D {\bf 72}, 053001 (2005)
  [arXiv:hep-ph/0507312];
%
S.~Choubey and W.~Rodejohann,
  Eur.\ Phys.\ J.\ C {\bf 40}, 259 (2005)
  [arXiv:hep-ph/0411190]; 
%
K.~Matsuda and H.~Nishiura,
Phys.\ Rev.\ D {\bf 72}, 033011 (2005)
[arXiv:hep-ph/0506192]; 
arXiv:hep-ph/0511338.


\bibitem{naturalness}
G.~'t Hooft,
PRINT-80-0083 (UTRECHT)
{\it Lecture given at Cargese Summer Inst., Cargese, France, Aug 26 - Sep 8, 1979}


\bibitem{seesaw} 
P.~Minkowski,
  Phys.\ Lett.\ B {\bf 67}, 421 (1977); 
T. Yanagida, in {\it Proc. of Workshop on Unified Theory and 
Baryon Number in the Universe}, eds. O. Sawada and A. Sugamoto, 
KEK, Tsukuba, (1979); 
%
M. Gell-Mann, P. Ramond and R. Slansky, in {\it Supergravity}, 
eds P. van Niewenhuizen and D. Z. Freedman (North Holland, Amsterdam 1980); 
P. Ramond, {\it  Sanibel talk}, retroprinted as hep-ph/9809459; 
%
S. L. Glashow, in {\it Quarks and Leptons}, Carg\`ese lectures, 
eds. M. L\'evy, (Plenum, New York, 1980) p. 707;  
%
R.~N.~Mohapatra and G.~Senjanovic,
  Phys.\ Rev.\ Lett.\  {\bf 44}, 912 (1980).
 

\bibitem{Mphase}
S.~M.~Bilenky, J.~Hosek and S.~T.~Petcov,
  Phys.\ Lett.\ B {\bf 94}, 495 (1980); 
J.~Schechter and J.~W.~F.~Valle,
  Phys.\ Rev.\ D {\bf 22}, 2227 (1980); 
M.~Doi, T.~Kotani, H.~Nishiura, K.~Okuda and E.~Takasugi,
  Phys.\ Lett.\ B {\bf 102}, 323 (1981).



\bibitem{yanagida}
T.~Yanagida, Talk at Japan-US Seminar on Double Beta Decay and Neutrino Mass, Maui, Hawaii, September 16-20, 2005. 


\bibitem{tHooft}
G.~'t Hooft,
  Phys.\ Rev.\ Lett.\  {\bf 37}, 8 (1976).

\bibitem{KRS} 
V.~A.~Kuzmin, V.~A.~Rubakov and M.~E.~Shaposhnikov,
  Phys.\ Lett.\ B {\bf 155}, 36 (1985).
  

\bibitem{coleman}
S.~Coleman, {\it Aspects of Symmetry : Selected Erice Lectures}
(Cambridge University Press, New York, 1988). 

 
\bibitem{manton}
  N.~S.~Manton,
  Phys.\ Rev.\ D {\bf 28}, 2019 (1983); 
F.~R.~Klinkhamer and N.~S.~Manton,
  Phys.\ Rev.\ D {\bf 30}, 2212 (1984).
  
  
\bibitem{mclerran}
  P.~Arnold and L.~D.~McLerran,
  Phys.\ Rev.\ D {\bf 37}, 1020 (1988); 
  Phys.\ Rev.\ D {\bf 36}, 581 (1987).


\bibitem {higher-dim}
S.~Weinberg,
  Phys.\ Rev.\ Lett.\  {\bf 43}, 1566 (1979).
F.~Wilczek and A.~Zee,
  Phys.\ Rev.\ Lett.\  {\bf 43}, 1571 (1979).
  

\bibitem {SKatm} 
Y.~Fukuda {\it et al.}  [Kamiokande Collaboration],
Phys.\ Lett.\ B {\bf 335}, 237 (1994);
%
Y.~Fukuda {\it et al.}  [Super-Kamiokande Collaboration],
Phys.\ Rev.\ Lett.\  {\bf 81}, 1562 (1998)
[arXiv:hep-ex/9807003];
Y.~Ashie {\it et al.}  [Super-Kamiokande Collaboration],
Phys.\ Rev.\ Lett.\  {\bf 93} (2004) 101801
[arXiv:hep-ex/0404034];
Y.~Ashie {\it et al.}  [Super-Kamiokande Collaboration],
  Phys.\ Rev.\ D {\bf 71}, 112005 (2005)
  [arXiv:hep-ex/0501064].


\bibitem{KamLAND}
K.~Eguchi {\it et al.} [KamLAND Collaboration],
Phys.\ Rev.\ Lett.\  {\bf 90}, 021802 (2003) 
[arXiv:hep-ex/0212021]; 
T. ~Araki {\it et al.} [KamLAND Collaboration],
Phys.\ Rev.\ Lett.\  {\bf 94}, 081801 (2005) 
[arXiv:hep-ex/0406035].


\bibitem {valle}
J.~W.~F. Valle, 
Talk at NO-VE 2006, in these Proceedings. 


\bibitem {K2K}
M.~H.~Ahn {\it et al.}  [K2K Collaboration],
Phys.\ Rev.\ Lett.\  {\bf 90}, 041801 (2003) 
[arXiv:hep-ex/0212007];
E.~Aliu {\it et al.}  [K2K Collaboration],
Phys.\ Rev.\ Lett.\  {\bf 94}, 081802 (2005) 
[arXiv:hep-ex/0411038].


\bibitem {solar}
B.~T.~Cleveland {\it et al.},
Astrophys.\ J.\  {\bf 496}, 505 (1998);
%
J.~N.~Abdurashitov {\it et al.}  [SAGE Collaboration],
Phys.\ Rev.\ C {\bf 60}, 055801 (1999)
[arXiv:astro-ph/9907113];
%
W.~Hampel {\it et al.}  [GALLEX Collaboration],
Phys.\ Lett.\ B {\bf 447}, 127 (1999);
%
S.~Fukuda {\it et al.}  [Super-Kamiokande Collaboration],
  Phys.\ Lett.\ B {\bf 539}, 179 (2002)
  [arXiv:hep-ex/0205075];
%
M.~B.~Smy {\it et al.}  [Super-Kamiokande Collaboration],
  Phys.\ Rev.\ D {\bf 69}, 011104 (2004)
  [arXiv:hep-ex/0309011];
%
Q.~R.~Ahmad {\it et al.}  [SNO Collaboration],
Phys.\ Rev.\ Lett.\  {\bf 87}, 071301 (2001)
[arXiv:nucl-ex/0106015];
{\it ibid.} {\bf 89}, 011301 (2002)
[arXiv:nucl-ex/0204008]; 
%
B.~Aharmim {\it et al.}  [SNO Collaboration],
  Phys.\ Rev.\ C {\bf 72}, 055502 (2005)
  [arXiv:nucl-ex/0502021].
  

\bibitem {davis} 
R.~Davis,
  Phys.\ Rev.\ Lett.\  {\bf 12}, 303 (1964).

\bibitem{SNO}
For SNO, see the last references in \cite{solar}.


\bibitem {QLC}
M.~Raidal,
  Phys.\ Rev.\ Lett.\  {\bf 93}, 161801 (2004)
  [arXiv:hep-ph/0404046];
%
H.~Minakata and A.~Y.~Smirnov,
  Phys.\ Rev.\ D {\bf 70}, 073009 (2004)
  [arXiv:hep-ph/0405088]. \\
%
For further references, see e.g., 
H.~Minakata,
  arXiv:hep-ph/0505262.


\bibitem {JPARC}
Y.~Itow {\it et al.}, arXiv:hep-ex/0106019.\\
For an updated version, see:
http://neutrino.kek.jp/jhfnu/loi/loi.v2.030528.pdf


\bibitem {NOVA}
D.~Ayres {\it et al.}  [Nova Collaboration],
  arXiv:hep-ex/0503053. 
 
 
\bibitem {OPERA}
M.~Komatsu, P.~Migliozzi, and F.~Terranova, 
J. Phys. G  {\bf 29}, 443 (2003) 
[arXiv:hep-ph/0210043].


\bibitem {krasnoyarsk}
Y.~Kozlov, L.~Mikaelyan and V.~Sinev,
Phys.\ Atom.\ Nucl.\  {\bf 66}, 469 (2003)
[Yad.\ Fiz.\  {\bf 66}, 497 (2003)]
[arXiv:hep-ph/0109277].


\bibitem {MSYIS} 
H.~Minakata, H.~Sugiyama, O.~Yasuda, K.~Inoue and F.~Suekane,
Phys.\ Rev.\ D {\bf 68}, 033017 (2003)
[Erratum-ibid.\ D {\bf 70}, 059901 (2004)]
[arXiv:hep-ph/0211111].

\bibitem{reactor_white}
K.~Anderson  {\it et al.},
arXiv:hep-ex/0402041. 


\bibitem{astro13}
 P.~D.~Serpico and M.~Kachelriess,
  Phys.\ Rev.\ Lett.\  {\bf 94}, 211102 (2005)
  [arXiv:hep-ph/0502088]; 
J.~F.~Beacom, N.~F.~Bell, D.~Hooper, S.~Pakvasa and T.~J.~Weiler,
  Phys.\ Rev.\ D {\bf 69}, 017303 (2004)
  [arXiv:hep-ph/0309267].
 

\bibitem{raghavan}
R.~S.~Raghavan,
  arXiv:hep-ph/0511191; 
  arXiv:hep-ph/0601079.


\bibitem {mikaelyan}
L.~A.~Mikaelyan, B.~G.~Tsinoev, and A.~A.~Borovoi, 
Sov. J. Nucl. Phys. {\bf 6}, 254 (1968) 
[Yad.\ Fiz.\  {\bf 6}, 349 (1967)]. 


\bibitem{visscher}
W.~M.~Visscher, 
Phys.\ Rev. {\bf 116}, 1581 (1959). 

\bibitem{schiffer}
W.~P.~Kells and J.~P.~Schiffer, 
Phys.\ Rev.\ C {\bf 28}, 2162 (1983). 


\bibitem {bahcall}
J.~N.~Bahcall, 
Phys.\ Rev.\  {\bf 124}, 495 (1961). 


\bibitem{mina-uchi}
H.~Minakata and S.~Uchinami,
  arXiv:hep-ph/0602046.
  

\bibitem{NPZ}
  H.~Nunokawa, S.~Parke and R.~Zukanovich Funchal, 
  Phys.\ Rev.\ D {\bf 72}, 013009 (2005)
  [arXiv:hep-ph/0503283].
  
\bibitem{GJK}
 A.~de Gouvea, J.~Jenkins and B.~Kayser,
  Phys.\ Rev.\ D {\bf 71}, 113009 (2005)
  [arXiv:hep-ph/0503079].
  

\bibitem{octant}
G.~Fogli and E.~Lisi, Phys.\ Rev.\ {\bf D54}, 3667 (1996);
[arXiv:hep-ph/9604415].


\bibitem{MSS04}
  H.~Minakata, M.~Sonoyama and H.~Sugiyama,
  Phys.\ Rev.\ D {\bf 70}, 113012 (2004)
  [arXiv:hep-ph/0406073].


\bibitem {resolve23}
 K.~Hiraide, H.~Minakata, T.~Nakaya, H.~Nunokawa, H.~Sugiyama, W.~J.~C.~Teves and R.~Z.~Funchal,
  arXiv:hep-ph/0601258.
  

\bibitem{concha-smi_23}
O.~L.~G.~Peres and A.~Y.~Smirnov,
  Phys.\ Lett.\ B {\bf 456}, 204 (1999)
  [arXiv:hep-ph/9902312];
%
  Nucl.\ Phys.\ B {\bf 680}, 479 (2004)
  [arXiv:hep-ph/0309312];
%
 M.~C.~Gonzalez-Garcia, M.~Maltoni and A.~Y.~Smirnov,
  Phys.\ Rev.\ D {\bf 70}, 093005 (2004)
  [arXiv:hep-ph/0408170].
  
  
\bibitem{choubey2}
S.~Choubey and P.~Roy,
  Phys.\ Rev.\ D {\bf 73}, 013006 (2006)
  [arXiv:hep-ph/0509197].


\bibitem{atm23}
T.~Kajita, 
Talk at Next Generation of Nucleon Decay and 
Neutrino Detectors (NNN05), 
Aussois, Savoie, France, April 7-9, 2005. 
http://nnn05.in2p3.fr/



  
\bibitem{MNjhep01}
H.~Minakata and H.~Nunokawa,
JHEP {\bf 0110}, 001 (2001) [arXiv:hep-ph/0108085];
Nucl.\ Phys.\ Proc.\ Suppl.\  {\bf 110}, 404 (2002) 
[arXiv:hep-ph/0111131].


\bibitem {MNplb97}
H.~Minakata and H.~Nunokawa,
Phys.\ Lett.\ B {\bf 413}, 369 (1997)
[arXiv:hep-ph/9706281].


\bibitem{intrinsic}
J.~Burguet-Castell, M.~B.~Gavela, J.~J.~Gomez-Cadenas, 
P.~Hernandez and O.~Mena,
Nucl.\ Phys.\ B {\bf 608}, 301 (2001)
[arXiv:hep-ph/0103258].


\bibitem{MNP2}
H.~Minakata, H.~Nunokawa and S.~Parke,
Phys.\ Rev.\ D {\bf 66}, 093012 (2002)
[arXiv:hep-ph/0208163].


\bibitem{T2KK}
  M.~Ishitsuka, T.~Kajita, H.~Minakata and H.~Nunokawa,
  Phys.\ Rev.\ D {\bf 72}, 033003 (2005)
  [arXiv:hep-ph/0504026].


\bibitem{kajita}
T.~Kajita, 
Talk at NO-VE 2006, in these Proceedings. 


\bibitem {superbeam}
H.~Minakata and H.~Nunokawa, Phys.\ Lett.\ {\bf B495} (2000) 369;
[arXiv:hep-ph/0004114];
J.~Sato, Nucl.\ Instrum.\ Meth.\ {\bf A472} (2001) 434 
[arXiv:hep-ph/0008056];
B.~Richter, arXiv:hep-ph/0008222.



\end{thebibliography}
\end{document}